# Self-Powered Wireless Sensor

Paul Ryuji Chuen-Ying Huang, Sing Kiong Nguang[1] and Ashton Partridge

Department of Electrical and Computer Engineering
University of Auckland, Auckland, New Zealand

**Abstract**

This paper develops a novel power harvesting system to harvest ambient RF energy to power a wireless sensor. Harvesting ambient RF energy is a very difficult task as the power levels are extremely weak. Simulation results show zero threshold MOSFETs are essential in the RF to DC conversion process. $0.5V_{DC}$ at the output of the RF to DC conversion stage is the minimum voltage which must be achieved for the micro-power sensor circuitry to operate. The weakest power level the proposed system can successfully harvest is -37dBm. The measured available power from the FM band has been measured to fluctuate between -33 to -43dBm using a ribbon FM dipole antenna. Ambient RF energy would best be used in conjunction with other forms of harvested ambient energy to increase diversity and dependability. The potential economic and environmental benefits make such endeavors truly worthwhile.

---

[1] Corresponding author, sk.nguang@auckland.ac.nz

# 1. Introduction

The contamination of landfills by batteries is a worldwide problem. Nearly three billion dry-cell batteries are consumed each year in the United States alone [1]. At the end of their operating life most batteries end in landfills. This contributes to a significant amount of hazardous compounds and heavy metals in the environment. To reduce reliance on batteries, energy harvesting techniques will be of vital importance in the near future. Energy harvesting is defined as the conversion of ambient energy into usable electrical energy [2-16].

The Citizen Eco-Drive is a famous example of energy harvesting. The watch collects ambient light and stores it as electrical energy in cells. These cells never need replacement and last longer than the life of the user [10], thus eliminating batteries ending up in landfills. There are many sources of ambient energy that can be harvested to perform work, such as solar power, wind power, energy from physical vibrations [12-15], thermal [11, 16], ocean tides and the RF spectrum. The primary focus of this investigation is to harvest energy directly from the RF spectrum. The RF spectrum covers the range from 3Hz to 300GHz [17]. The frequency band of interest is from the FM transmission band, it is located between 87.5 and 108 MHz of the RF spectrum [18]. There are two main reasons for selecting FM, firstly because of the extensive coverage and the polarization of the electro-magnetic fields are circular, which will enable the energy harvesting antenna to be positioned in virtually any orientation and still be able to receive a signal. The energy collected will be used to power a wireless temperature and humidity sensor node. The wireless and battery-less capability of these devices will greatly increase their utility.

The ultimate aim is to be capable of harvesting energy from across the entire RF spectrum, to maximize the amount of power collected. Additional investigation has been conducted, to further understanding on the challenges of harvesting RF energy at higher frequency levels. The frequency investigated is the GSM 900 cellular network. Historically, the attempts at harvesting energy from the cellular frequency band have only been of limited success as one is either restricted to transmitting their own power [19-20] or limited by the distance from the actual transmission tower [21].

This paper is structured as follows. Section 2 introduces the circuitry of the proposed RF power harvesting system. Section 3 provides an overview of the energy management system. Section 4 presents the energy harvesting performance of the prototype and also provides discussions on the results and the implications. Section 5 discusses future work is required. This paper is concluded in Section 6.

**2. Energy Harvesting Circuitry**

The proposed energy harvesting topology is shown in Figure 1.

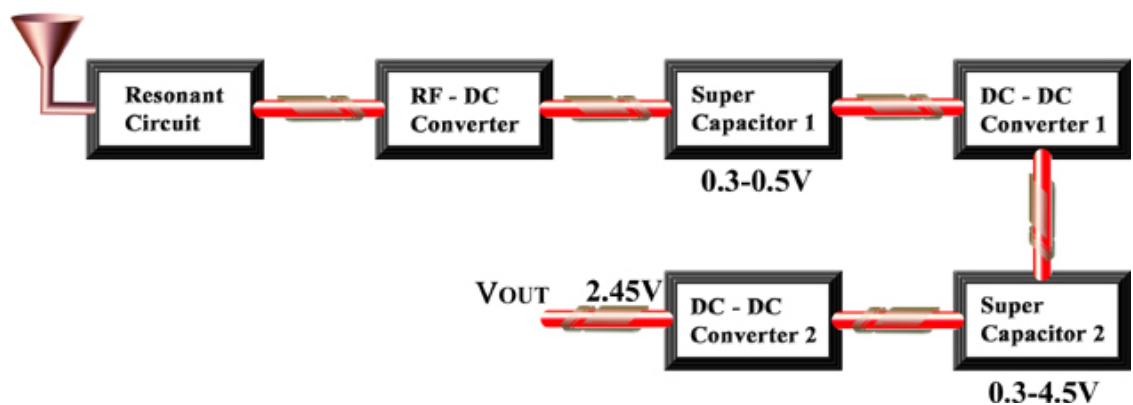

**Figure 1**: RF energy harvesting sequence

An impedance matching network has not been implemented between the energy harvesting antenna and the resonant circuit. This is because the energy harvesting circuit is a non-linear load, the input impedance changes with changes in voltage. Since the antenna and the input is not matched, reflection will always occur. The tuned series resonant circuit boosts the voltage amplitude of the incoming RF signal. The RF-DC converter rectifies the zero mean RF signal into DC and stores the charge on Super Capacitor 1. Super Capacitor 1 is used as a temporary storage capacitor, once the voltage reaches 0.5V, DC – DC Converter 1 transfers the energy to Super Capacitor 2 where the energy is stored at a higher voltage. This enables much more energy to be stored. DC-DC converter 2 is used to supply a constant 2.45V to the digital electronic components of the sensor node. By using a second DC-DC converter the voltage across Super Capacitor 2 can drop down to 0.3V before the sensor node will stop functioning.

*2.1. Resonant Circuit*

To increase the output voltage amplitude, the inductance is made as large as possible when compared with the capacitance value as shown in Figure 2. Surface mount components are used to reduce parasitic capacitance. The actual components are selected by their frequency response performance. The capacitors used are Class I ceramic capacitors. Their capacitance can be assumed to be frequency independent up to 1GHz [22]. The inductors used are high frequency multilayer chip inductors. The inductors are made from advanced ceramics with low resistance silver as the conductor [23]. The physical layout on the PCB is also important. To prevent the PCB from appearing as a transmission line, the components are placed as close together as possible. To reduce mutual inductance between inductors, the inductors are placed at 90 degrees to each other. For harvesting at FM transmission frequencies the resonant frequency is set at 100MHz.

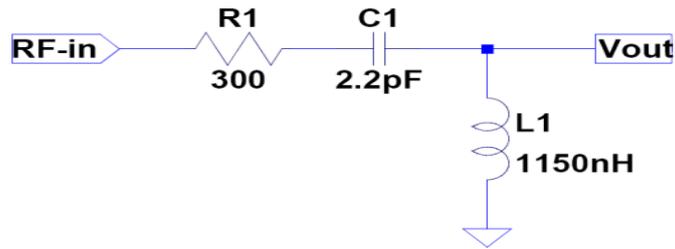

**Figure 2**: Series tuned resonant circuit

*2.2. RF-DC Converter*

The RF-DC converter is realized by cascading voltage doublers into a large voltage multiplier. In total 25 stages have been implemented. The Zero threshold MOSFETs are operating in the cut-off region, therefore the weak inversion current is utilized to rectify the RF signal as shown in Figure 3. Zero threshold MOSFETs reduce the voltage drop across each stage, therefore it increases the number of stages which can be cascaded. This is effectively Self-Synchronous Rectification (SSR).

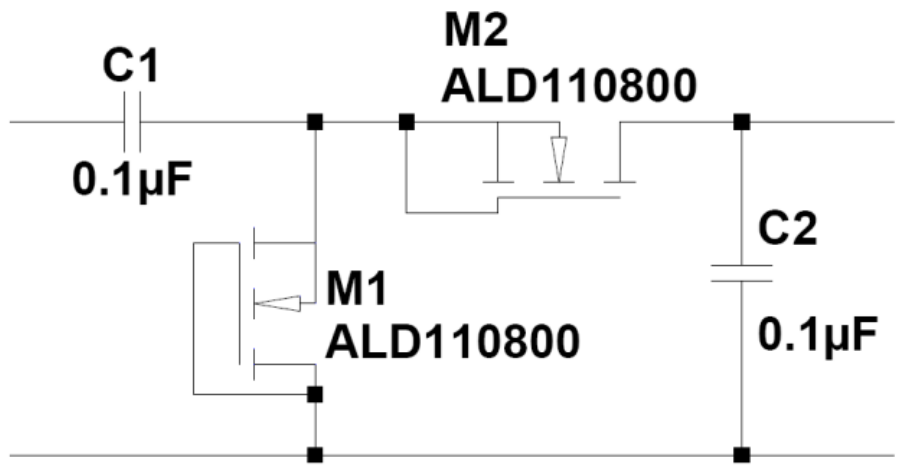

**Figure 3**: One voltage doubling stage

The capacitors used in the RF-DC converter are NP0 Class I ceramic capacitors. These are selected because of their excellent frequency response and the low leakage characteristics of the dielectric material. These are both very important qualities as the voltage multiplier is designed to work at frequencies in the MHz range. Also in order to improve efficiency any leakage current must be minimized.

*2.3. Super Capacitor*

Super capacitors are selected over conventional energy storage devices such as lithium-ion or lithium-ion polymer cells, because super capacitors have a close to infinite charging and discharging life cycles. There are two types of super capacitors used in the energy harvesting stage. The first super capacitor is used as a temporary energy storage device. Therefore an ordinary 1.5 Farad super capacitor is used to lower the cost of implementation. The second super capacitor is used to store the collected energy at a higher voltage to maximize the amount of energy which can be stored as shown by Equation 1.

$$E = \frac{1}{2}CV^2 \qquad (1)$$

The second super capacitor is a 1 Farad Bestcap, it has an ultra-low ESR, together with an extremely low leakage current [9]. A low leakage current is critical because harvesting ambient RF energy is very difficult therefore any charge collected is very valuable.

*2.4. DC-DC Converter*

The micro-power energy harvesting circuitry design contains two DC-DC converters. Existing available integrated chip converters were chosen over designing a converter from scratch because of several key reasons. The first reason is because of the very high efficiencies which can be achieved by using existing integrated converters. Efficiencies up to 90% can be easily attained. The second reason is a DC-DC converter which is able to accept a very large input voltage range is required. The input voltage range can vary between 0.3 to 4.5V. It is very difficult to manually implement a boost converter capable of operating with an input voltage of 0.3V. The final reason is integrated converters can automatically switch between buck and boost conversion thus simplifying development. The DC-DC converter selected for this application is the TPS61200 from Texas Instruments. It has a minimum startup voltage of 0.5V and can operate down to 0.3V. Both DC-DC converters are the same to simplify the design and manufacturing process.

**3. Energy Management System**

In Figure 4, the arrows show the direction of data flow. Some data lines are bi-directional. In Appendix B the schematic for the entire energy management system and the digital circuitry can be located.

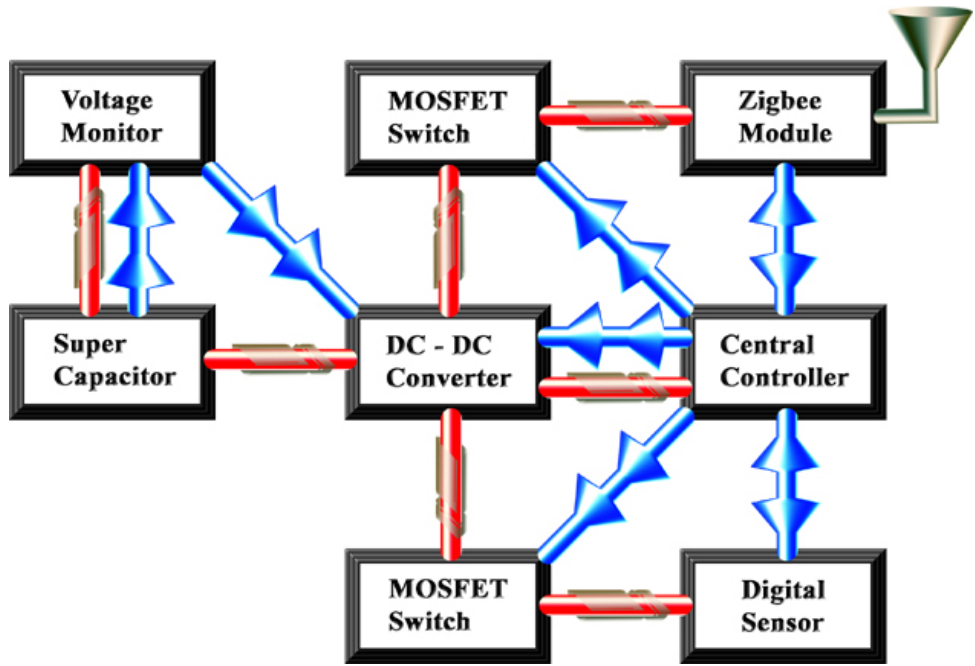

**Figure 4:** Sensor power and data flow

*3.1. Voltage Monitor*

The heart of the energy management system is the voltage monitor. It is implemented using an ATmega165P microcontroller. The voltage monitor operates at a clock frequency of 32.768 kHz and is powered directly from the super capacitor as shown in Figure 7. The voltage at which the voltage monitor starts to operate is 1.8V. The voltage monitor consumes 0.6 uA of current while in power-save mode. While in power-save mode, the real time clock is still operating. Once every week the voltage monitor will wake up from the power-save mode and check the voltage across the super capacitor. The regularity at which the voltage monitor checks the voltage stored on the super capacitor can be programmed to suit the particular application. Once sufficient voltage has been harvested, the voltage monitor enables the DC-DC converter which enables the central controller to operate. The voltage is checked through the ADC of the microcontroller. The code for operating the ADC is located in Appendix C. The ADC requires a voltage reference. The internal voltage reference has been utilized to save on external components and can be turned on and off to conserve power.

*3.2. Central Controller energy management*

The central controller is also implemented using an ATmega165P microcontroller. It was selected because of its extremely low power consumption of a mere 10 uA in the active mode. Once the central controller powers up, the first priority is to gain control of the DC-DC converter. This is because the voltage monitor will return back to power-save mode to conserve energy. Without control of the DC-DC converter the central controller will have no power source. To gain control of the DC-DC converter the central controller must set the I/O pin which is connected to the enable pin of the DC-DC converter to high. The central controller is capable of monitoring the voltage across the super capacitor. This enables the controller to know how much time is left for the tasks to be completed. Once the data has been transmitted back to the data recorder, the central controller will disable the DC-DC converter.

*3.3. Power MOSFET switch*

Although both the zigbee module and the digital sensor have sleep modes, they still consume current. To eliminate this unnecessary leakage current, MOSFET switches have been added to allow the central controller full control of when each device will receive power. This enables precise timing to be achieved thus minimizing the amount of power used. The MOSFET used are the IRLML2502PbF from International Rectifier. These are selected because it has a maximum threshold voltage of 1.2V, this is essential as the operating voltage within the sensor node is 2.45V.

Another advantage with the IRLML2502PbF MOSFET it has a very low ON resistance of 0.045 ohms. Both the zigbee module and the digital sensor node have a MOSFET switch connected.

## 4. Results and Discussions

The power levels were measured using a spectrum analyzer. Two antennas have been tested at FM transmission levels. The first antenna is a length of insulated copper wire which has been cut to half a wavelength, to resemble a monopole antenna. A SMA connector is used to connect this antenna to the spectrum analyzer. The amount of power measured at the output of the antenna is -50dBm. The second antenna is a standard ribbon FM dipole antenna. The power measured at the output varied between -33dBm and -43dBm. This antenna is shown in Figure 5.

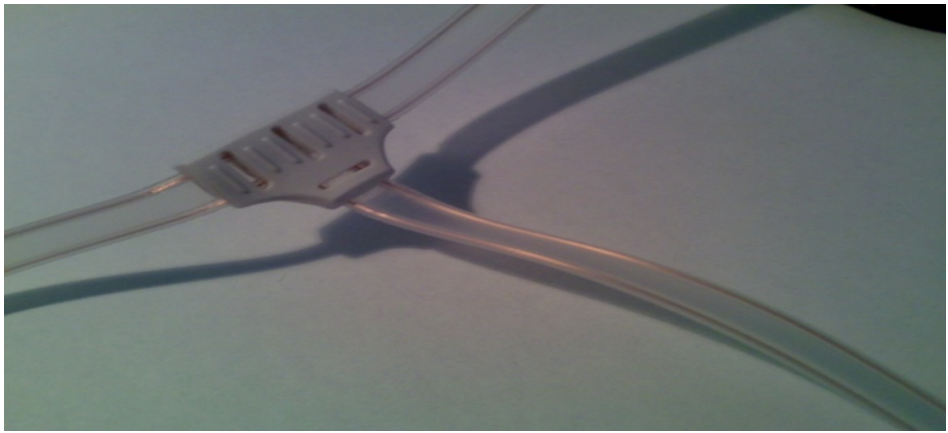

**Figure 5**: Ribbon FM dipole antenna

The procedure used to test the RF energy harvesting performance of a given circuit is as follows. The circuit input is connected to the output of a precision signal generator. The signal generators are capable of varying the amount of power in dBm and also the signal frequency. $0.5V_{DC}$ at the output of the RF to DC conversion stage is the minimum voltage which must be achieved for the micro-power sensor circuitry to operate. $0.5V_{DC}$ is therefore the standard by which we compare the two different rectifying schemes.

A total of 20 cascaded voltage doubling stages have been built to test the performance of Schottky diodes in rectifying small signals. The smallest power that could be rectified to 0.5V was -18dBm. It is also worthy to note after 7 to 8 doubling stages have been cascaded, the voltage gain of adding additional stages diminishes. To further improve the energy harvesting circuit, a series resonant circuit was added to boost the amplitude of the incoming RF signal. The hardware implementation was successful. A 9 dB voltage gain has been achieved. However this is at the cost of compromising the RF bandwidth from which we are harvesting. The top and bottom layers of the prototype are shown in Figures 6 and 7, respectively.

The energy required for each device is based on how long each device is operating and the amount of current each device draws. Power calculations for a typical wireless sensor application are tabulated in Table 1. It is shown that the energy required to complete one data transmission cycle is 0.32 Joules. The result has been confirmed by measuring the voltage difference before and after data transmission.

| Device | V (V) | I (mA) | T (s) | E (J) |
|---|---|---|---|---|
| ATmega165P | 1.8 | 0.01 | 10 | 0.00018 |
| ATmega165P | 1.8 | 0.01 | 8 | 0.00014 |
| Sensor | 3.3 | 0.55 | 5 | 0.00908 |
| Zigbee | 3.3 | 35 | 2.7 | 0.31 |
| **Total** | | | | **0.32** |

**Table 1**: Power Calculations for a wireless sensor

Assuming there is a constant -37dBm power flow entering the energy harvesting antenna. It will take 11.7 days for the RF harvester to collect 0.32 Joules of energy. This does not take into account the reflection which may occur between the antenna and the input of the RF harvester,

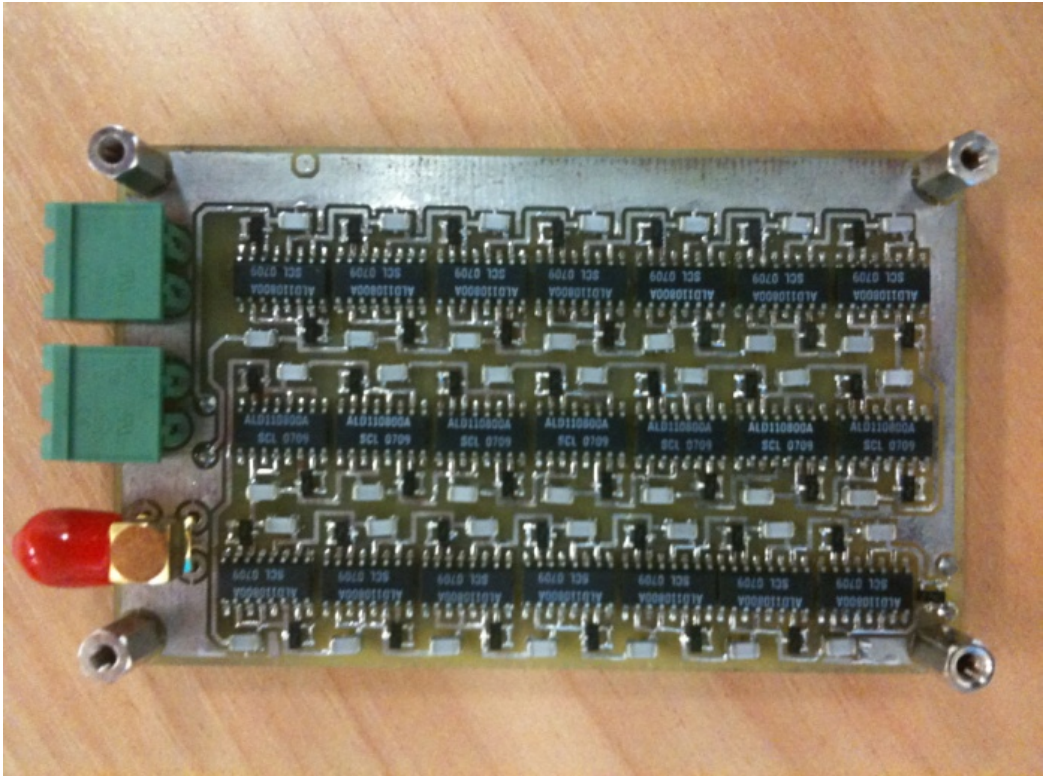

**Figure 6**: Power harvesting system (top layer)

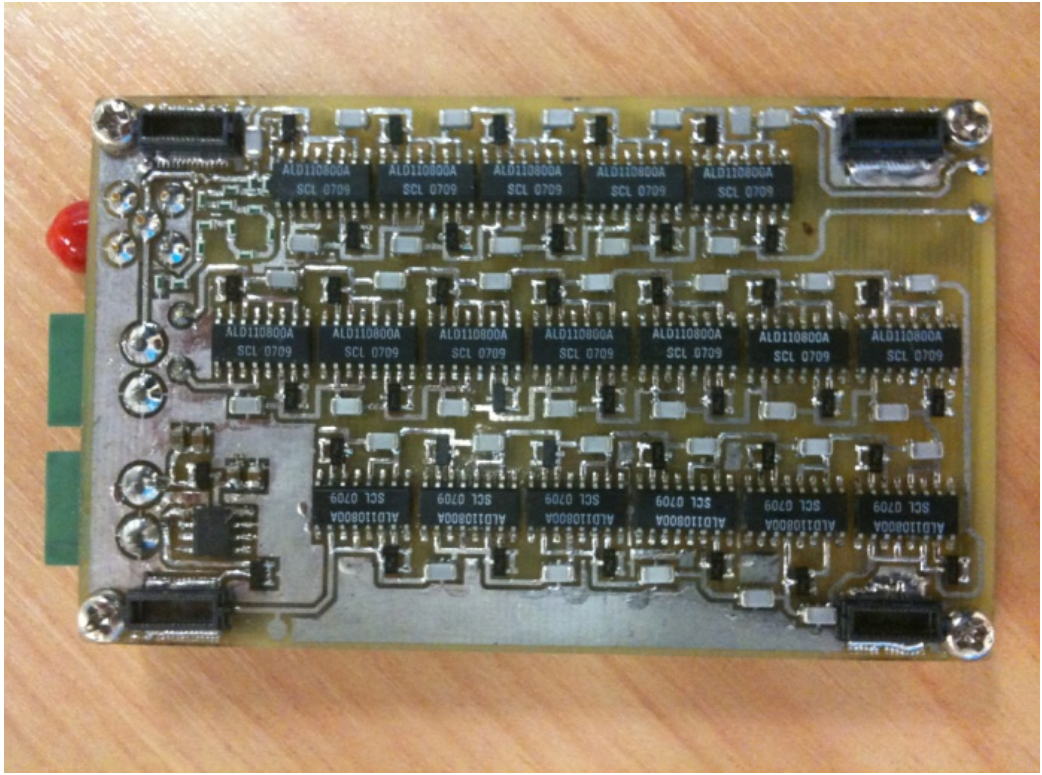

**Figure 7**: Power harvesting system (bottom layer)

The realistic time the RF energy harvesting circuit will take to collect enough energy for one data transmission will be closer to 20 to 30 days. This is taking into account the reflection coefficient and the fluctuations in power levels. There are still many applications where such a device will be very useful. One such application is in silver-culture. Here circumferential sensors are used to measure the growth of trees. Since the trees grow very slowly the time between measurements is usually a month or more. By powering these sensors from an ambient source, it eliminates the need to ever change the batteries. For applications requiring faster charge rates, it is possible to combine energy harvested from various different ambient sources to increase the available power and diversity.

## 5. Future Developments

To be truly successful at harvesting ambient RF energy, the entire energy harvesting circuitry must be integrated on a single IC. This is to reduce leakage losses, increases the overall efficiency of the system and help energy harvesting from the RF spectrum to be economically viable. For this project the objective was to harvest only from the FM transmission band, however the power collected is very limited and the time taken to harvest the energy is very long. Therefore the next step is to increase the bandwidth from which the RF energy harvester can collect energy. This is no trivial task, because in radio systems it is easy design a circuit to work at a certain frequency but for the circuit to work at multiple frequencies may require different harvesting techniques. It will be necessary to design a broadband antenna. Another area which needs improvement is the power sensitivity. It is highly probable that by making the entire circuitry physically smaller the system will be capable of harvesting even lower power levels. A possible future development is to integrate RF energy harvesting with other forms of ambient energy such heat and physical vibrations.

## 6. Conclusions

Harvesting ambient RF energy is very difficult as the power levels are very weak. Schottky diode based voltage multipliers are not suitable for rectifying such weak signals. A tuned series resonant circuit enables the energy harvesting circuitry to harvest at lower power levels but the bandwidth is compromised. The real hardware implementation supports the simulation results showing that SSR using n-MOSFETs is essential in the RF to DC conversion process. The lowest power level which can be boosted and rectified is -37dBm at 100MHz. At 900MHz the lowest power level the energy harvesting circuit can harvest increased to -25dBm therefore the harvesting performance at higher frequencies is not as good. Since the measured available power from a FM ribbon dipole antenna fluctuated between -33 and -43dBm, harvesting energy from the FM spectrum it feasible.